# Effect of chemical modification of edge atoms of graphene sheets on their strength

**V. A. Popova, N. A. Popova, E.F. Sheka**

Peoples' Friendship university of Russia, Moscow, 117198 Russia

Basing on the molecular theory of graphene, the influence of the chemical modification of edge atoms of a graphene sheet is studied in terms of mechanochemical reactions. The mechanical behavior of graphene is shown to be not only highly anisotropic with respect to the direction of the load application, but greatly dependent on the chemical modification of the edges as well. A topological character of the graphene deformation is discussed.

Oppositely to real physical experiments, when changing the object shape under loading is usually monitored, computational experiments deal with the total energy response to the object shape deformation that simulates either tension and contraction or bending, screwing, shift, and so forth. As for graphene, whose mechanical properties are amenable to experimental study with difficult, the computational experiments takes on great significance.

A lot of works are devoted to the calculation of mechanical properties of graphene due to which two approaches, namely, continuum and atomistic ones have been formulated. The continuum approach is based on the well developed theory of elasticity of continuous solid media applied to shells, plates, beams, rods, and trusses. The latter are structure elements used for the continuum description. When applying to graphene, its lattice structure is presented in terms of the above continuum structure elements and the main task of the calculation is the reformulation of the total energy of the studied atomic-molecular system subjected to changing in shape by that in terms of the continuum structure elements. This procedure involves actually the adaptation of the theory of elasticity of continuous media to nanosize objects which makes allowance for introducing macroscopic basic mechanical parameters such as Young's modulus ($E$), the Poisson ratio ($\nu$), the potential energy of the elastic deformation, etc into the description of mechanical properties of graphene. Since the energy of graphene is mainly calculated in the framework of quantum chemistry, which takes the object atom structure into account, the main problem of the continuum approach is a linkage between molecular configuration and continuum structure elements. Nanoscale continuum methods (see Refs. 1-5 and references therein), among which those based on the structural mechanics concept [6] are the most developed, have shown the best ability to simulate nanostructure materials. In view of this concept, graphene is a geometrical frame-like structure where the primary bonds between two nearest-neighboring atoms act like load-bearing beam members, whereas an individual atom acts as the joint of the related beams [7-10].

The basic concept of the atomistic approach consists in obtaining mechanical parameters of the object from results of the direct solutions of either Newton motion laws [10, 11] or Schrödinger equations [12, 13] under changing the object shape following a particular algorithm of simulation of the wished type of deformation. It should be necessary to issue a general comment concerning calculations based on the application of the DFT computational schemes. All the latter, except the recent one [14], were performed in the framework of restricted versions of the programs that do not take into account spins of the graphene odd electrons and thus ignore the correlation interaction between these electrons. The peculiarities of the graphene odd electron behavior are connected with a considerable enlarging of its C-C bonds, which, in its turn, causes a noticeable weakening of the odd electron interaction and thus requires taking into account these electrons correlation [15].

In the case of atomistic approach, not energy itself, but forces applied to atoms become the main goal of calculations. These forces are input later into the relations of macroscopic linear theory of elasticity and lay the foundation for the evaluation of micro-macroscopic mechanical parameters such as Young's modulus ($E^*$), the Poisson ratio ($\nu^*$), and so on. Nothing to mention that parameters $E$ and $E^*$ as well as $\nu$ and $\nu^*$ are not the same so that their coincidence is quite accidental. Obviously, atomistic approach falls in opinion comparing with the continuum one due to time consuming calculations and, as a result, due to applicability to smaller objects. However, it possesses doubtless advantages concerning the description of the mechanical behavior of the object under certain loading (shape changing) as well as exhibiting the deformation and failure process at atomic level.

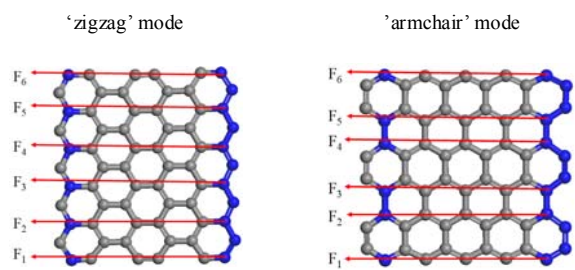

**Fig.1**. Six mechanochemical internal coordinates of uniaxial tension of the molecule (5,5) NGr for two deformation modes. $F_1$, $F_2$, $F_3$, $F_4$, $F_5$ и $F_6$ are forces of response along these coordinates. Blue atoms fix the coordinates ends.

Recently a new atomistic approach has been suggested for the description of the graphene deformation based on considering the failure and rupture process of graphene as the occurrence of a mechanochemical reaction [16-18]. A similarity between mechanically induced reaction and the first-type chemical ones, first pointed out by Tobolski and Eyring more than sixty years ago [19], suggested the use of a well developed quantum-chemical approach of the reaction coordinate [20] in the study of atomic structure transformation under deformation. Firstly applied to the deformation of poly(dimethylsiloxane) oligomers [21], the approach has revealed a high efficacy in disclosing the mechanism of failure and rupture of the considered polymers. It has been successfully applied recently for the description of uniaxial tension of graphene [16, 17] and graphane [18] molecules thus exhibiting itself as a significant part of the current molecular theory of graphene [22].

The main point of the approach concerns the reaction coordinate definition. When dealing with chemical reactions, the coordinate is usually selected among the internal ones (valence bond, bond angle or torsion angle) or is presented as a linear combination of the latter. Similarly, mechanochemical internal coordinates (MICs) are introduced as modified internal coordinates defined in such a way as to be able to specify the considered deformational modes [21, 23]. Thus, uniaxial tension and contraction are described by linear MICs similar to valence bonds. Figure 1 presents two sets of MICs that describe a tensile deformation of the (5, 5) NGr graphene molecule. The molecule has been led into the foundation of previously performed computational experiments [16-18] and presents a rectangular fragment of a graphene sheet cut along zigzag and armchair edges and containing 5 benzenoid units along each direction. The two MICs set are designed along and normally to the C-C bonds chains of the molecule for allowing the molecule deformation in two deformational modes, namely, armchair and zigzag ones, to be considered. In due course of calculations, the main idea of which is found the minimum of the total energy at each deformation step of a stepwise change of MICs, the MICs are excluded from the optimization procedure. This allows determining the force of response as the residual gradient of the total energy along the selected MIC. The results presented in paper were obtained in the framework of the Hartree-Fock unrestricted (UHF) version of the DYQUAMECH codes [24] exploiting advanced semiempirical QCh methods (PM3 version [25]).

The forces of response $F_i$ applied along the $i^{th}$ MICs are the first derivatives of the total energy $E(R)$ over the Cartesian coordinates [7]:

$$\frac{dE}{dR} = <\varphi\left|\frac{\partial H}{\partial R}\right|\varphi> + 2<\frac{\partial \varphi}{\partial R}|H|\varphi> + 2<\frac{\partial \varphi}{\partial P}|H|\varphi>\frac{dP}{dR} \quad (1)$$

Here $\varphi$ is the wave function of atom of the ground state at fixed nucleus positions, $H$ presents the adiabatic electron Hamiltonian, and $P$ is the nucleus momentum. When the force calculation is completed, the gradients are re-determined in the system of internal coordinates in order to proceed further in seeking the total energy minimum by atomic structure optimization. Forces $F_i$ are used afterwards for determining all required micro-macroscopic mechanical characteristics, which are relevant to uniaxial tension, such as the total force of response $F = \sum_i F_i$, stress

$\sigma = F/S = \left(\sum_i F_i\right)/S$, where $S$ is the loading area,

the Young's modulus $E = \sigma/\varepsilon$, where both stress $\sigma$ and the strain $\varepsilon$ are determined within the elastic region of deformation.

As was shown earlier [16-18], the elastic region of tensile deformation of both graphene and graphane (5, 5) NGr molecules is extremely narrow and corresponds to a few first steps of the deformation. The deformation as a whole is predominantly plastic and dependent on many parameters. Equilibrium structures of the molecule before and after uniaxial tension, which was terminated by the Equilibrium structures of the molecule before and after uniaxial tension, which was terminated by the rupture of the last C-C bond coupling two fragments of the molecule, are shown in Fig. 2. Looking at the picture lets us notice two main peculiarities of the molecule deformation. First concerns the anisotropy of the deformation with respect to two deformational modes. Second exhibits a strong dependence of the deformation on the chemical composition of the molecule edge atoms. The deformation anisotropy of graphene has been attributed to mechanical anisotropy of the constituent benzenoid units [16, 17]. The dependence of the deformation on the chemical modification of framing edge atoms has been revealed for the first time.

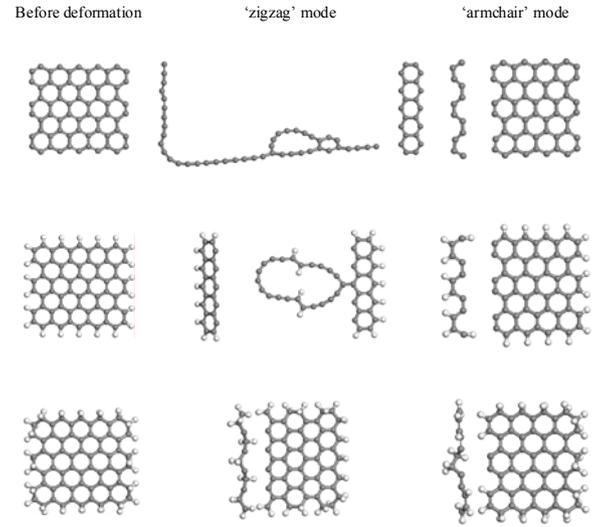

**Fig.2**. Equilibrium structures of the (5,5) NGr with different chemical modification of edge atoms before and after completing tensile deformation in two modes of deformation. Bare edges (top); $H_1$-terminated edges (middle); $H_2$-terminated edges (bottom).

As seen in Fig. 2, when the edge atoms are bare and not terminated by other ones, the deformation behavior is the most complex. The deformation development has a tricotage-like character which is why the deformational modes 'zigzag' and 'armchair' have many differences (see detail description in [16, 17]). In the former case, the deformation is multi-stage and consists of 250 consequent steps with elongation of 0.1Å at each step. In contrast, the deformational mode 'armchair' is one-stage and is terminated on the 17$^{th}$ step of the deformation.

**Table 1**. Young's modules for (5,5) NGr with different configuration of edge atoms, TPa

| Mode | Bare edges | $H_1$-terminated edges | $H_2$- terminated edges |
|---|---|---|---|
| 'zigzag' | 1.05 | 1.09 | 0.92 |
| 'armchair' | 1.06 | 1.15 | 0.95 |

The addition of one hydrogen atom to each of edge atoms of the molecule does not change the general character of the deformation: it remains a tricotage-like one so that there is still a large difference between the behavior of 'zigzag' and 'armchair' modes. At the same time, the number of the deformation steps of 'zigzag' mode reduces to 121.

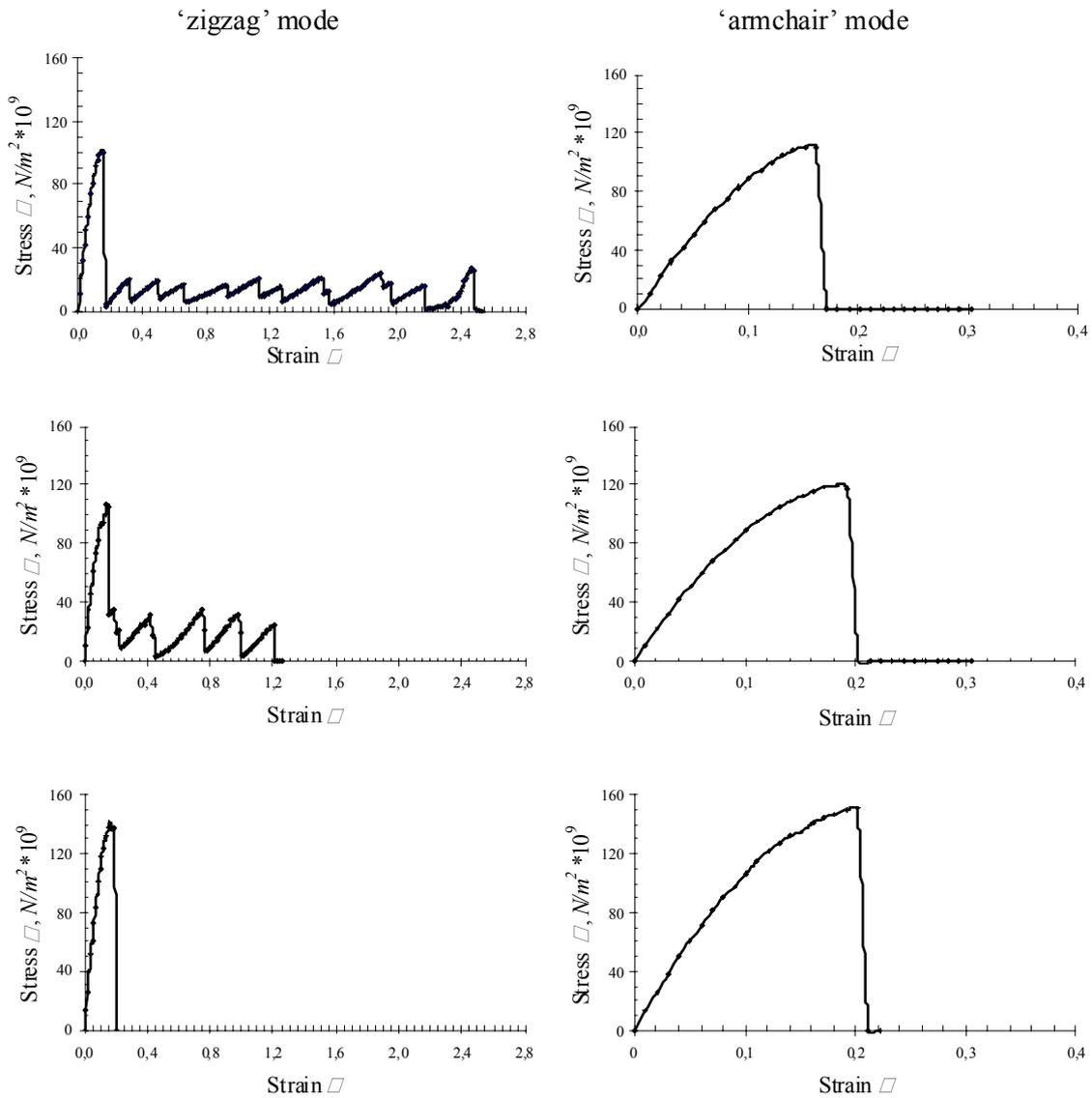

**Fig,3**. Stress-versus-strain dependences of tensile deformation of the (5, 5) NGr molecule with different chemical modification in two deformation modes. Bare edges (top); $H_1$-terminated edges (middle); $H_2$-terminated edges (bottom).

The addition of the second hydrogen atoms to the edge ones drastically changes the situation. The tricotage-like character disappears and both 'zigzag' and 'armchair' modes become one-stage and quite short by deformation steps.

Figure 3 presents a set 'stress-strain' relations that fairly well highlight the difference in the mechanical behavior of all the three molecules. Table 1 presents Young's modules that were defined in the region of the elastic deformation. As seen from the table, the Young's modules depend on the character of edge atom chemical modification. As shown in [18], elastic properties of big molecules such as polymers [21, 26] and nanographenes [18] are determined by dynamic characteristics of the objects, namely, by force constants of the related vibrations. Since benzenoid units have a determining resistance to any deformation of graphene molecules, the dynamic parameters of stretching C-C vibrations of the units are mainly responsible in the case of uniaxial tension. Changing in Young's modules means changing in the force constants (and, consequently, frequencies) of these vibrations. The latter are attributed to G-band of graphene that lays the foundation of a mandatory testing of any graphenium system by Raman spectrum. In many cases, the relevant band is quite wide which might indicate the chemical modification of the edge zone of the graphene objects under investigation.

Since the molecule deformation is mainly provided by basal atoms, changing in Young's modules points to a significant influence of chemical state of edge atoms on the electronic properties in basal plane. Obviously, it is resulted from a significant correlation of the molecule odd electrons which causes their conjugation. The correlation itself changes quite remarkably in due course of the deformation. This can be illustrated by the evolution of the total number of effectively unpaired electrons $N_D$ during the deformation. The $N_D$ value is a direct characteristic of the extent of the electron correlation, on one hand, [22] and molecular chemical susceptibility, on the other, [27].

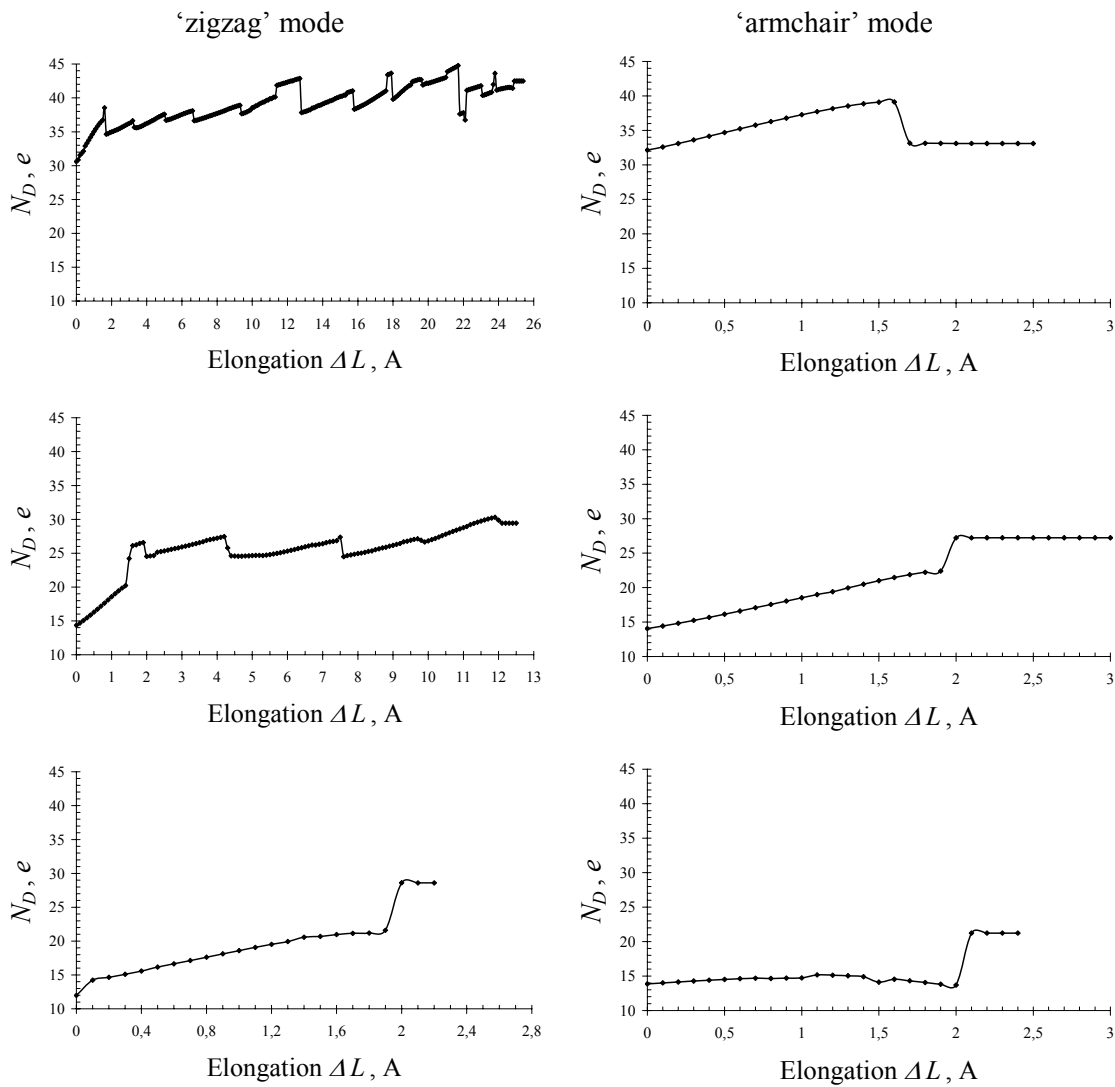

**Fig.4.** Evolution of the odd electrons correlation interms of the total numbers of effectively unpaired electrons under tensile deformation of the (5, 5) NGr at two deformation modes. Bare edges (top); $H_1$-terminated edges (middle); $H_2$-terminated edges (bottom).

Changing in $N_D$ in due course of deformation exhibits changing in the chemical activity of the molecule induced by deformation.

Figure 4 presents the evolution of $N_D$ for the three studied molecules. Since breaking of each C-C bonds causes an abrupt changing in $N_D$, a toothed character of the relevant dependences related to 'zigzag' mode of the molecule with bare and H-terminated edges is quite evident. One should draw attention to the $N_D$ absolute values as well as to their dependence on both chemical modification of the edge atoms and the deformational modes. Evidently, chemical activity of the molecules is drastically changed in due course of a mechanically induced transformation, This changing is provided by the redistribution of C-C bond lengths caused by the mechanical action.

Presented in the current paper undoubtedly shows that the chemical modification of the graphene molecule edge atoms has a great impact on its mechanical behavior. The feature is a result of a significant correlation of the molecule odd electrons followed by their conjugation over the molecule. Thus, the transition from the molecule with bare edges with maximal correlation of odd electrons to the molecule with $H_1$- and $H_2$-terminated edges is followed with a considerable supressing of the correlation related to the edge atoms in the former case and with a complete zeroing of the latter in the latter case. It turns out, that the changes are not local and strongly influence the electronic structure in the region of the basal plane, where the main deformational process occurs. Such a strong influence of the edge atoms evidently tells us as well about a topological character of the deformational process in graphene.